\documentclass[%
 reprint,
nofootinbib,
 amsmath,amssymb,
 aps,
]{revtex4-2}

\usepackage{graphicx}
\usepackage{dcolumn}
\usepackage{bm}
\usepackage{mathtools}
\usepackage{lipsum}
\usepackage{titlesec}
\makeatletter
\newcommand{\two}{\bBigg@{2.5}}
\newcommand{\three}{\bBigg@{3.5}}
\newcommand{\four}{\bBigg@{4.5}}
\newcommand{\five}{\bBigg@{5.5}}
\newcommand{\seven}{\bBigg@{7.5}}
\newcommand{\fourteen}{\bBigg@{17}}
\makeatother

\titlespacing\subsection{0pt}{10 pt plus 0 pt minus 0 pt}{10 pt plus 0 pt minus 0 pt}
\titlespacing\subsubsection{0pt}{9 pt plus 1 pt minus 0 pt}{9 pt plus 1 pt minus 0 pt}
\usepackage{hyperref}

\begin{document}

\preprint{APS/123-QED}

\title{Using the Climate App to learn about\\ Planetary Habitability and Climate Change}

\author{Lan Xi Zhu}

 \email{lan.zhu@mail.mcgill.ca}
 
 \author{Anthony Courchesne}%

 \email{courchea@mila.quebec}
 
 \author{Nicolas B.\ Cowan}%

 \email{nicolas.cowan@mcgill.ca
}

\affiliation{%
$^*\, ^\ddagger$Department of Earth \& Planetary Sciences and Department of Physics\\McGill University, Montréal, QC, Canada\\$^\dagger$School of Computer Science, McGill University and \\Montreal Institute for Learning Algorithms\\Montréal, QC, Canada
}%

\date{\today}

\begin{abstract}
Simple climate models have been around for more than a century but have recently come back into fashion: they are useful for explaining global warming and the habitability of extrasolar planets.  The Climate App (\href{https://www.climateapp.ca/}{www.climateapp.ca}) is an interactive web-based application that describes the radiative transfer governing planetary climate. The App is currently available in French and English and is suitable for teaching high-school through college students, or public outreach. The beginner version can be used to explore the greenhouse effect and planetary albedo, sufficient for explaining anthropogenic climate change, the Faint Young Sun Paradox, the habitability of TRAPPIST planets and other simple scenarios. There is also an advanced option with more atmospheric layers and incorporating the absorption and scattering of shortwave radiation for students and educators wishing a deeper dive into atmospheric radiative transfer. A number of pedagogical activities are being beta tested and rolled out. 
\end{abstract}

\maketitle


\section{\label{sec:intro} Introduction}
Global warming has become a matter of increasing  concern in recent decades \cite{matzner2019astronomy,stevens2020imperative}. In parallel, astronomers have discovered thousands of exoplanets with varied atmospheres, some of which could harbour life. Despite tremendous public and academic interest in both topics, it is not widely appreciated that the same 19$^{\rm th}$ century atmospheric physics governing anthropogenic climate change also governs planetary habitability \cite{foote1856art,arrhenius1896xxxi}. 

A visually intuitive and interactive way to convey the concepts of albedo and the greenhouse effect can be useful for students, educators and the public, e.g.,  \cite{Singh,Caballero}.  This motivated the web-based Climate App, which serves as a tool to explore the roles that atmospheric absorption and emission play in planetary climate. 
We introduce each component of the Climate App in Section \ref{sec:UI} and explain the underlying physical concepts and mathematical  model in Section \ref{sec:physics}.

\section{\label{sec:UI} The User Interface}
\begin{figure*}[htb]
    \centering
    \includegraphics[width=0.7\textwidth]{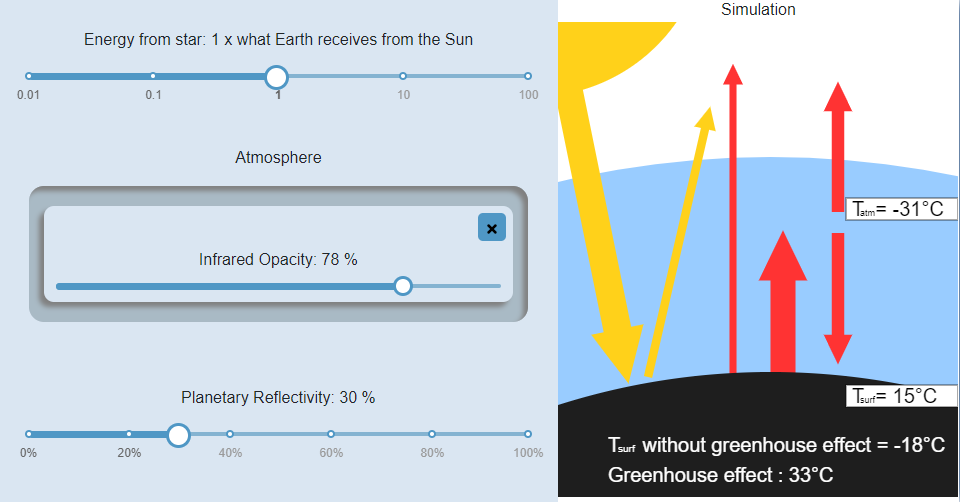}
    \caption{Screenshot of the Beginner version of the Climate App.}
    \label{fig:diagrams}
\end{figure*}

The Climate App is a single-page web application
written in HTML5, javascript and CSS using React.js v16.5.1 so it works on a computer, tablet or smartphone. The app is currently available in French and English. You can change between the beginner and advanced version using the switch below the title. The two versions differ in the complexity of their underlying atmospheric models, and consequently have different number of adjustable parameters (sliders). As shown in Figure \ref{fig:diagrams}, the user interface consists of a control panel and a graphical panel that changes in real time as the parameters are adjusted.
\subsection{\label{sec:Trenberth} The Trenberth Diagram}
The graphical panel of the Climate App is a visual representation of the energy flows controlling the planet's climate. The pale blue swath above the planet's surface represents the atmosphere. The yellow arrows show shortwave radiation from the star, roughly corresponding to visible light for the Sun. The red arrows represent longwave thermal emission, typically infrared, from the planetary surface and atmosphere. 
 The width of each arrow is proportional to the importance of that energy flow as compared to the amount of starlight that reaches the planet. This sort of schematic showing energy flows in a planet's atmosphere is sometimes called a Trenberth Diagram and are commonplace in climate studies of Earth and other worlds \cite{2009BAMS...90..311T,2016QJRMS.142..703R,2019ApJ...884L...2S}.

Moving the sliders in the control panel changes the planetary parameters and hence the way that energy flows in and out of the atmosphere, which is ultimately what governs the surface temperature. Temperatures are reported in Celsius for the beginner version and Kelvin for the advanced version. The beginner version of the app also reports the magnitude of the greenhouse effect in degrees Celsius; this is the difference between the surface temperature and the value it would take if the infrared opacity of the atmosphere were zero. The default values of the sliders when you open the Climate App, both the beginner and advanced versions, are approximately the average values for modern-day Earth.

\subsection{\label{sec:Sliders} The Sliders}The control panel on the left lets you vary the planet's properties and hence its climate with the corresponding sliders. Three properties are common to both versions while two extra properties are present exclusively in the Advanced version.\\

\noindent\emph{Energy from Star (Stellar Radiation in Advanced App)}\\
Also known as ``instellation'', this slider controls the amount of starlight that reaches the top of the planet's atmosphere. For planets in the Solar System, the star in question is none other than the Sun!  Incident starlight is the most important factor determining a planet's climate. For convenience, the Climate App expresses the instellation in terms of the solar constant, the average flux that the Earth receives from the Sun (341 W/m$^2$).  
For example, a value of 10 on this slider would mean that the planet receives ten times the amount of stellar radiation that we receive on Earth. The slider is a logarithmic scale, which allows for values a hundred times smaller or greater than the solar constant.\\

\noindent\emph{Planetary Reflectivity (Surface Albedo in Advanced App)}\\
This slider controls the surface reflectivity of the planet. 
A value of 0 means the planet is perfectly black and absorbs all incoming stellar radiation; there are many planets close to this limit. No planet, on the other hand, reflects \textit{all} incoming radiation, so the Climate App restricts the albedo to values less than 99\%. 

Albedo and instellation determine how much energy the planet absorbs every second and hence how much energy it must radiate away every second in order to remain in radiative equilibrium.  Although this energy balance is only approximate from one second to another, it must be maintained on timescales of years.\footnote{Secular climate change may lead to slight radiative imbalances, e.g., under anthropogenic global warming Earth radiates away slightly less power than it absorbs from the Sun---the extra power primarily warms the ocean.} 

In the advanced app, starlight can also be reflected back to space by the atmospheric layer(s) so the top-of-atmosphere effective albedo is reported in the Trenberth diagram.  The effective albedo can be greater than or less than the underlying surface albedo, depending on the shortwave properties of the atmosphere.  

The default value for the planetary reflectivity in the beginner app is 30\%, while in the advanced app the surface albedo is 18\% and the effective albedo---accounting for scattering by the atmosphere---is 30\%. The Earth's albedo is approximately 30\%, largely due to clouds. When this slider is moved, the planet in the simulation canvas change shade.\\

\noindent\emph{Atmosphere (Atmospheric Layers in Advanced App)}\\
This box contains properties that are specific to the planetary atmosphere. In the beginner version, the atmosphere is treated as a single layer (Figure \ref{fig:diagrams}). You may remove the atmosphere by clicking the $\times$ and add it back by clicking on a + box that only appears when your planet does not have an atmosphere.

\begin{figure}[htp]
    \centering
    \includegraphics[width=0.38\textwidth]{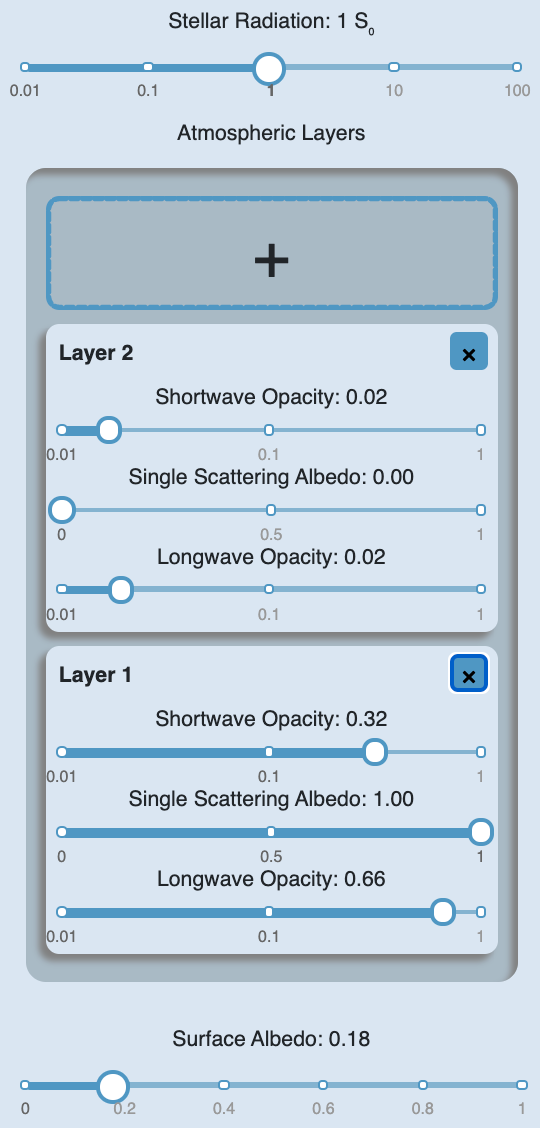}
    \includegraphics[width=0.38\textwidth]{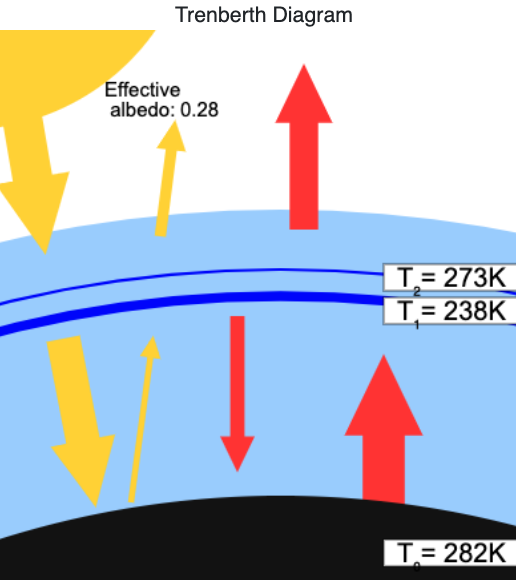}
    \caption{Screenshot of the Advanced version of the Climate App, as it appears on a smartphone.}
    \label{fig:adv}
\end{figure}

For the advanced version (Figure \ref{fig:adv}), a maximum of three atmospheric layers are allowed and each can be assigned different properties. This simulates vertical changes in atmospheric properties, e.g., Earth's atmosphere has more water vapour in the lower, warmer layers compared to the higher and colder layers. In this case, it is more realistic to assign greater infrared opacity to lower layers instead of setting one single value representing the whole atmosphere. You can click on the + box to add a new layer, and the $\times$ on a layer to remove it.\\

\noindent \emph{Infrared Opacity (Longwave Emissivity in Adv.\ App)}\\ Each atmospheric layer has an adjustable longwave emissivity, its effectiveness in absorbing and emitting thermal radiation. A value of 1 absorbs all infrared radiation that hits it and re-radiates that energy equally up and down. This value depends on the greenhouse gases (GHG) in the atmosphere: for small levels of GHG the infrared opacity increases linearly with gas abundance, but once absorption lines are saturated---as they are on Earth---opacity increases more slowly.\\ 
    
\noindent \emph{Shortwave Opacity (only in Advanced app)}\\ As shortwave radiation encounters an atmospheric layer, the photons may pass through without interaction, be absorbed, or be scattered by molecules and aerosols in the atmosphere. This slider determines the fraction of shortwave radiation interacting with the corresponding layer (absorbed or scattered). A value of 1 means that all of the shortwave photons are either absorbed or scattered by that atmospheric layer.\\
    
\noindent\emph{Single Scattering Albedo (only in Advanced app)}\\ Single scattering albedo refers to the fraction of interacting shortwave photons that are scattered rather than absorbed. For example, a value of 0.1 indicates that out of all the shortwave photons having interacted with the atmosphere, 10\% are scattered, while the remaining 90\% are absorbed. The Climate App assumes isotropic scattering, so half of the scattered photons keep going in their original direction of travel, while the other half are back-scattered from whence they came.

\section{\label{sec:physics}The Underlying Physics}
\subsection{Key Concepts}
The Climate App is built upon atmospheric physics that have been known since the late 19$^\mathrm{th}$ century\cite{arrhenius1896xxxi}. The following key concepts are essential for users or students who wish to dive deeper into the model derivation.
\subsubsection{Shortwave and Longwave Radiation}
All objects emit radiation, whether you can see it or not. The colour-dependence of this so-called blackbody radiation is quantitatively described by Planck's Law. Hot objects like the Sun emit most of their light at visible wavelengths (the range of light that human eye can see, not coincidentally). Somewhat cooler objects like planets and people tend to emit most of their radiation at longer infrared wavelengths (beyond the red end of the rainbow).  Since visible and thermal light interact with the planetary surface and atmosphere in qualitatively different ways, it is customary in climate science to distinguish between the shortwave radiation received from a star and longwave radiation emitted by the planet (among experts, this is known as the two band approximation). In the Climate App visualizations, shortwave light is denoted by yellow arrows, while thermal radiation is denoted by red arrows.

\subsubsection{The Stefan-Boltzmann Law}

The Stefan-Boltzmann law quantifies the relation between the temperature and total emitted flux (at all wavelengths) for a perfectly absorbing and emitting object, a so-called blackbody: $F_{\rm bb}=\sigma T^4$,
where $F_{\rm bb}$ is the emitted flux (units of W/m$^2$), $T$ is the temperature of the object in Kelvin, and $\sigma$ is the Stefan-Boltzmann constant:
$\sigma=5.67 \times 10^{-8}$ W m$^{-2}$ K$^{-4}$.

Real materials are not perfect blackbodies: they do not perfectly absorb and radiate.  Since the surface and atmosphere of a planet emit at longer, thermal wavelengths, we quantify the ability of an object to radiate by its longwave emissivity, $\epsilon_{\rm LW}$. The flux emitted by an element of the climate system is therefore given by $F=\epsilon_{\rm LW}\sigma T^4$.  The surface emits this amount of radiation upwards, while an atmospheric layer emits this amount of radiation upwards and the same amount downwards.

\subsubsection{Kirchhoff's Law of Thermal Radiation}
Kirchhoff's law of thermal radiation dictates that a material's ability to emit radiation at some wavelength is equal to its ability to absorb radiation at that same wavelength.    For example, gases that readily absorb infrared light---so-called greenhouse gases---must also emit well at those same wavelengths.

Solid and liquid surfaces clearly do not absorb all of the light that hits them (otherwise they would appear perfectly black!).  Indeed, the surface reflectance (or albedo) slider in the Climate App controls the fraction of visible light that the planetary surface reflects.\footnote{In principle, surfaces imperfectly radiate visible light, but planetary surfaces and atmospheres are rarely hot enough to radiate appreciably in the visible, so the Climate App implicitly neglects shortwave thermal emission.}  These same materials, however, are very close to being perfect blackbodies in the thermal infrared. They all have $\epsilon_{\rm LW} \approx 1$. To very good approximation, we therefore adopt  $\epsilon_{\rm LW} \equiv 1$ for the planetary surface in the Climate App.

A planet's atmosphere, on the other hand, can have an infrared emissivity anywhere between zero to one ($0\le\epsilon_{\rm LW}\le1$), depending on the presence of greenhouse gases; this is precisely what the \emph{Longwave Opacity} slider controls (\emph{Infrared Opacity} in the beginner version). In the beginner version the atmosphere cannot absorb shortwave radiation, so if the longwave emissivity is set to zero then the atmosphere simply cannot absorb any energy and its temperature is ill-defined and we remove the atmospheric temperature label.\footnote{In a real planetary atmosphere even transparent layers can be heated via conduction or convection, but these processes are neglected in the Climate App.}

In the beginner version of the app we assume that shortwave radiation does not interact with the atmosphere ($\epsilon_{\rm SW} \equiv 0$). This is approximately correct for planets like Earth and Mars, where most Sunlight indeed makes it down to the surface of the planet, but is utterly inadequate for Venus or Titan. Such thick atmospheres require the Advanced Climate App, which allows for the absorption or scattering of shortwave radiation ($0<\epsilon_{\rm SW}\le1$).\subsubsection{Radiative Equilibrium}
A planet must radiate as much energy to space as it absorbs from its star, at least on average.  Moreover, if radiation is the only way to move energy around in a climate system, then the total absorbed and emitted flux must be equal for each component of the climate system (atmospheric layer or planetary surface). To understand why this equilibrium arises, imagine a patch of ground that absorbs more energy than it emits: it will warm up and hence radiate away more heat following the Stefan-Boltzmann Law, finding a new, warmer, equilibrium. Conversely, a region that radiates more energy than it absorbs will cool down.\footnote{The Climate App is inherently a one-dimensional model of the atmosphere: it represents global mean properties of a planet.  As such, it neglects the transport of heat from one region to another via winds or ocean currents.}

Radiation is not the only way to move energy around. Heat conduction is rarely important for planetary climate\footnote{A notable exception is understanding the seasonal cycles under ground, useful for planting gardens and building foundations.} so we completely neglect it in the Climate App. Convection, on the other hand, operates in many planetary atmosphere, including the Earth's. Since convection is just another means of moving heat from the planetary surface up to the atmosphere, it can roughly be thought of as contributing to the red arrow pointing from the surface up to the atmosphere.


The concept of radiative equilibrium is illustrated in the right panel of the Climate App: the total width of all arrows from the planetary surface equal the total width of all arrows reaching the surface. The same situation applies for each atmospheric layer, and indeed even at the top of the atmosphere: the same amount of energy enters the atmosphere from above as exits to space.

\subsubsection{The Two Stream Approximation}
Photons can go in any direction, but it is expedient for one-dimensional climate calculations such as these to assume that light is either travelling straight up or straight down (this simplification is known to aficionados as the two stream approximation).  Conceptually, it is easy to see how the messy reality of multi-directional photons might boil down to the two-stream approximation: any ray of light must be travelling somewhat towards or away from the planet, but typical rays must pass through a bit more atmosphere because they are not entering or exiting the atmosphere perfectly vertically.

\subsection{Beginner App: Single-Layer Model}
\begin{figure}[htb]
    \centering
    \includegraphics[width=0.4\textwidth]{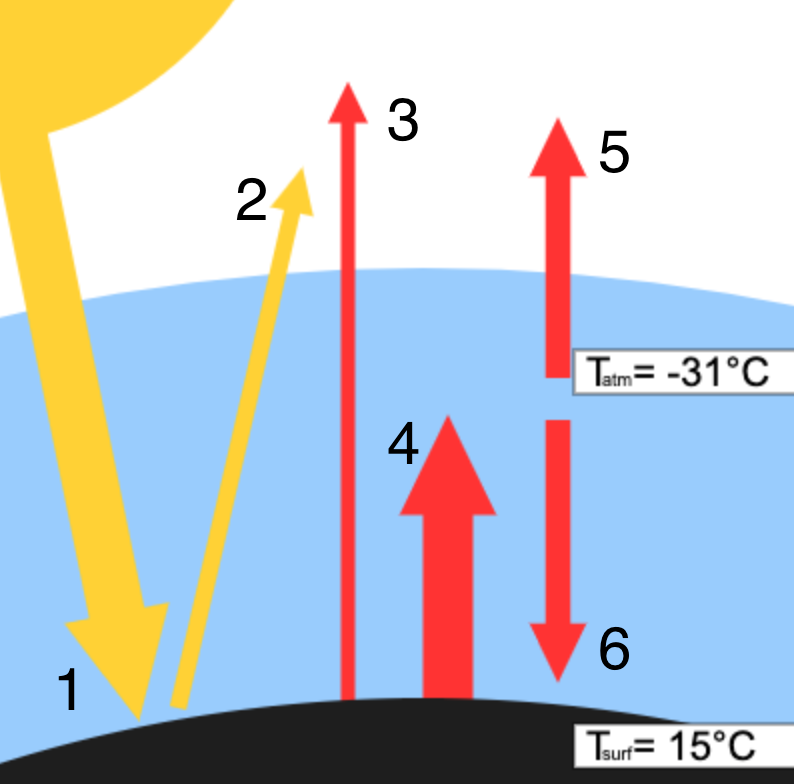}
    \caption{Schematic showing radiative fluxes in the beginner version of the Climate App.}
    \label{fig:arrows}
\end{figure}

In the Beginner version we consider the simplest greenhouse model, where the incoming solar radiation passes through the atmosphere without incident. The flow of energy under this simple model is illustrated in Figure \ref{fig:arrows}. 
Incoming solar radiation (1) is partially reflected back into space (2). The absorbed power heats up the surface, which radiates away energy as longwave radiation (red arrows). Some of this energy makes it to space (3) while some is absorbed by the atmosphere (4). The atmosphere radiates away the energy it absorbs to remain in radiative equilibrium. Unlike the surface, the atmosphere radiates both upwards (5) and downwards (6). 

Since the temperatures of the surface and the atmosphere are interdependent, we can establish a system of equations to solve for both temperatures. In fact, a more complex model with three or more layers follows the same principle, but the system of equations becomes more involved, as shown below. 

The guiding principle is radiative equilibrium: each component of the climate system must emit as much flux as it absorbs. There are two unknowns: the surface temperature, $T_{\rm surf}$, and the atmospheric temperature, $T_{\rm atm}$.  We therefore need two equations in order to solve the problem. In each case we put the absorbed power on the left hand side and the emitted power on the right. The atmospheric balance is 
\begin{equation}\label{eq:surf}
\epsilon_{\rm LW} \sigma T_{\rm surf}^4 = 2\epsilon_{\rm LW} \sigma T_{\rm atm}^4,
\end{equation}
where $\epsilon_{\rm LW}$ is the longwave emissivity (infrared opacity in the Beginner version). The left hand side is the flux emitted by the surface that is absorbed by the atmosphere (arrow 4) and the right hand side is the sum of the upward and downward fluxes emitted by the atmosphere (arrows 5 and 6). Inspection of Equation \ref{eq:surf} reveals many of the same symbols on both sides of the equality, which begs for simplification, yielding $T_{\rm surf}^4 = 2 T_{\rm atm}^4$.

The flux balance at the surface is
\begin{equation}
F_* - A F_* + \epsilon_{\rm LW} \sigma T_{\rm atm}^4 = \sigma T_{\rm surf}^4,
\end{equation}
where $F_*$ is the stellar flux reaching the planet\footnote{Climate scientists often denote this by $S_0/4$, where $S_0$ is the ``Solar constant'' and the factor of four dilution accounts for the ratio between the cross-sectional area and surface area of a planet.} and $A$ is the planetary albedo (planetary reflectivity in the Beginner App). The first term on the left hand side is the incident stellar flux (arrow 1), the second term is the reflected light (arrow 2) and the third term is the downward atmospheric emission (arrow 6); on the right hand side there is the emission from the planetary surface (arrow 4)---recall that we assume perfect surface emissivity, an excellent approximation.

Substituting our expression for atmospheric radiative equilibrium, $T_{\rm surf}^4 = 2 T_{\rm atm}^4$, into Equation \ref{eq:surf}, we obtain
\begin{equation}
(1-A)F_* + \epsilon_{\rm LW} \sigma T_{\rm atm}^4 = \sigma (2T_{\rm atm}^4),
\end{equation}
which can be solved for the atmospheric temperature:
\begin{equation}
T_{\rm atm} = \left(\frac{F_*}{\sigma}\right)^{1/4}\left( \frac{1-A}{2-\epsilon_{\rm LW}}\right)^{1/4}
\end{equation}
and from thence the surface temperature:
\begin{equation}
T_{\rm surf} = \left(\frac{F_*}{\sigma}\right)^{1/4}\left( \frac{1-A}{1-\epsilon_{\rm LW}/2}\right)^{1/4}.
\end{equation}

For the Climate App's default values of $A=0.30$ and $\epsilon_{\rm LW}=0.78$, we obtain $T_{\rm atm} = 242$ K $= -31^\circ$C and $T_{\rm surf} = 288$ K $= 15^\circ$C, as shown in Fig. \ref{fig:diagrams} and in good agreement with average values on Earth today.

Lastly, the Beginner version of the Climate App reports the magnitude of the greenhouse effect.  This is calculated by comparing the surface temperature with what it would have been in the absence of the greenhouse effect, i.e., $\epsilon_{\rm LW}=0$.  For the default parameters, the magnitude of the greenhouse effect is 33$^\circ$C, again in good agreement with modern-day Earth.

\subsection{Advanced App: Three-Layer Model}
Using the same reasoning as the simple greenhouse model, we can derive the system of equations to solve for the temperatures of the planetary surface and three atmospheric layers.  However, the model used in the advanced version is more complicated. In addition to the extra atmospheric layers, the advanced Climate App includes two new parameters for each atmospheric layer to control how they interact with shortwave radiation.  These parameters are the shortwave opacity, $\epsilon_{\rm SW}$, and the single-scattering albedo, $\alpha$.  

\subsubsection{Shortwave Radiation}
When shortwave radiation arrives at an atmospheric layer, it can do three things: transmission with a factor of $1-\epsilon_{\rm SW}$, scattering with a factor of $\epsilon_{\rm SW}\alpha$, and absorption with a factor of $\epsilon_{\rm SW}(1-\alpha)$. We use $F_3$, $F_2$, $F_1$ and $F_0$ to denote the shortwave flux absorbed at layer 3, layer 2, layer 1 of the atmosphere and at the surface, respectively. 

The values of the shortwave flux terms $F_3$, $F_2$, $F_1$ and $F_0$ are determined as follows. In the one-dimensional scattering model used in the Climate App, we assume an asymmetry factor of 0 at each atmospheric layer, i.e., isotropic scattering. This means that half of the photons  scattered at a given layer are scattered back in the direction they came from, while the other half are scattered forward. Since we use a 1-D model, we treat the forward scattering case as equivalent to transmission. With the isotropic scattering assumption, we can solve the following system of equations to obtain the shortwave terms:
\begin{align*}
    F_3 &= (I+u_2)\epsilon_{\rm SW3}(1-\alpha_3)\\
F_2 &= (d_3+u_1)\epsilon_{\rm SW2}(1-\alpha_2)\\
F_1 &= (d_2+u_0)\epsilon_{\rm SW1}(1-\alpha_1)\\
F_0 &= d_1(1-A)\\
u_0&= d_1A\\
d_1&= 0.5u_0\epsilon_{\rm SW1}\alpha_1+d_2(1-\epsilon_{\rm SW1})+0.5d_2\epsilon_{\rm SW1}\alpha_1\\
u_1&= u_0(1-\epsilon_{\rm SW1})+0.5d_2\epsilon_{\rm SW1}\alpha_1+0.5u_0\epsilon_{\rm SW1}\alpha_1\\
d_2 &= 0.5u_1\epsilon_{\rm SW2}\alpha_2+d_3(1-\epsilon_{\rm SW2})+0.5d_3\epsilon_{\rm SW2}\alpha_2\\
u_2 &= 0.5d_3\epsilon_{\rm SW2}\alpha_2+u_1(1-\epsilon_{\rm SW2})+0.5u_1\epsilon_{\rm SW2}\alpha_2\\
d_3&= 0.5u_2\epsilon_{\rm SW3}\alpha_3+I(1-\epsilon_{\rm SW3})+0.5I\epsilon_{\rm SW3}\alpha_3\\
u_3 &= u_2(1-\epsilon_{\rm SW3})+0.5I\epsilon_{\rm SW3}\alpha_3+0.5u_2\epsilon_{\rm SW3}\alpha_3,
\end{align*}
where $I$ is the incident stellar flux, $u_i$ and $d_i$ are the total shortwave radiation heading upward and downward from the $i^{\rm th}$ layer, $\epsilon_{\rm SW1}$, $\epsilon_{\rm SW2}$ and $\epsilon_{\rm SW3}$ are the shortwave opacities of each atmospheric layer, $\alpha_1$, $\alpha_2$ and $\alpha_3$ are their single scattering albedo, and $A$ is the surface albedo. 
\vspace{0.2cm}
\begin{figure}[htb]
    \centering
    \includegraphics[width=0.45\textwidth]{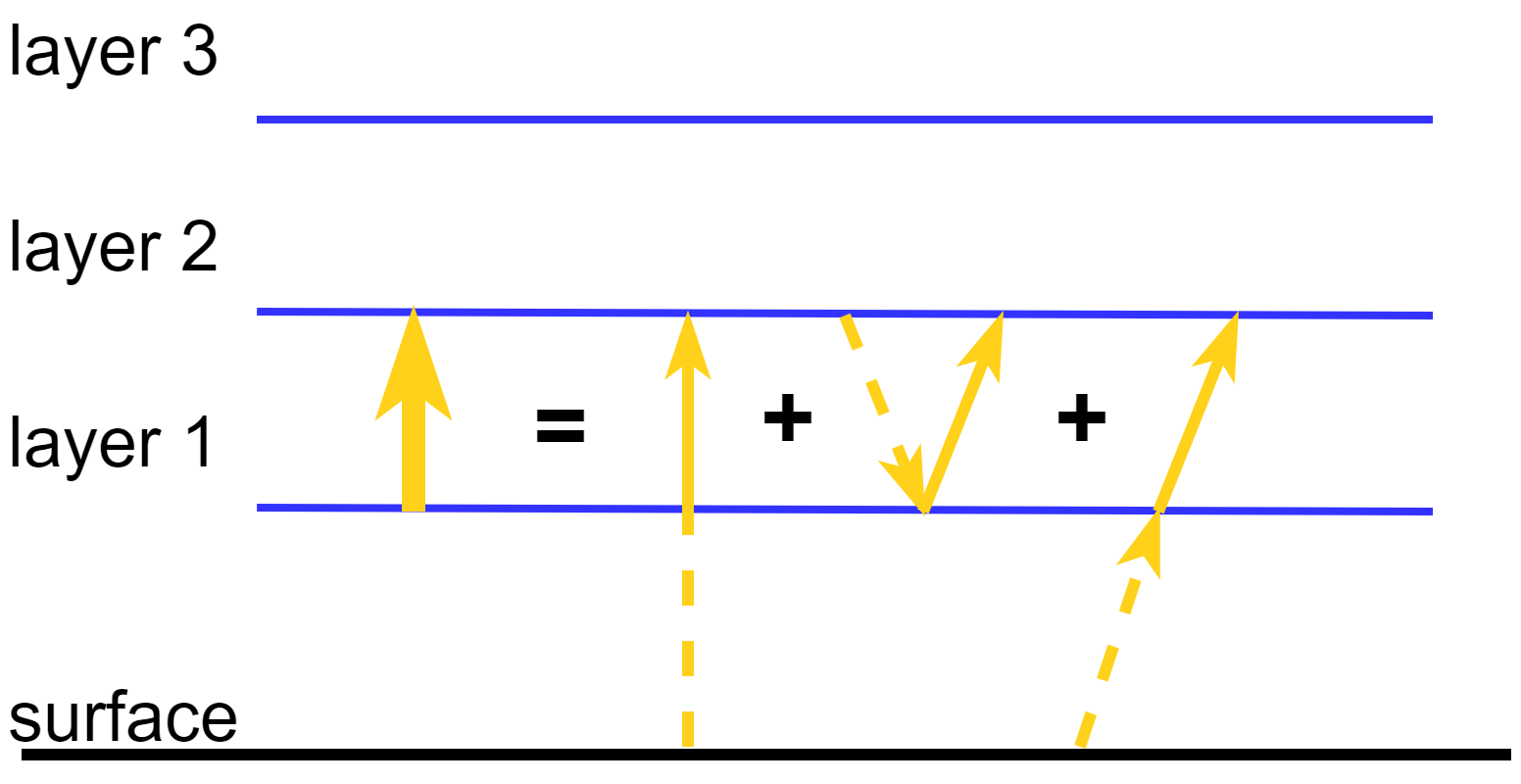}
    \caption{Diagram showing how the expression of $u_1$ (the leftmost thick arrow) was obtained.}
    \label{fig:scatter}
\end{figure}
As an example, Figure \ref{fig:scatter} shows the shortwave radiation transmitted upward from layer 1, $u_1$. It is the sum of three components indicated by three thinner arrows (from left to right): a fraction of shortwave radiation reflected upward from the surface that goes through layer 1 without any interaction, $u_0(1-\epsilon_{\rm SW1})$, a fraction of radiation transmitted downward from layer 2 that back-scatters off of layer 1, $\frac{1}{2}d_2\epsilon_{\rm SW1}\alpha_1$, and another fraction of radiation reflected upward from the surface that forward scatters off layer 1, $\frac{1}{2}u_0\epsilon_{\rm SW1}\alpha_1$. The factors of $\frac{1}{2}$ comes from the assumption of isotropic scattering.

\subsubsection{Longwave Radiation}
Following the logic that led us to Equation \ref{eq:surf}, we focus on layer 2 (with absorbed shortwave flux $F_2$, temperature $T_2$ and longwave emissivity $\epsilon_{\rm LW2}$) as an example to the derivation of the full system of equations shown in Appendix \ref{ap:3layer}. The emission from layer 2 is $2\epsilon_{\rm LW2} \sigma T_2^4$, identical to the right hand side of Equation \ref{eq:surf} since layer 2 also emits both upwards and downwards. As for the absorption, we have shortwave radiation $F_2$, thermal radiation from layer 3 and layer 1, and attenuated thermal radiation from the surface. The equation of radiative equilibrium is therefore:
\begin{align}\label{eq:layer2}
    2\epsilon_{\rm LW2} \sigma T_2^4 = & F_2 + \sigma T_0^4\epsilon_{\rm LW2}(1-\epsilon_{\rm LW1}) \nonumber\\
    &+ \sigma T_1^4\epsilon_{\rm LW1}\epsilon_{\rm LW2}+ \sigma T_3^4\epsilon_{\rm LW2}\epsilon_{\rm LW3}.
\end{align}
The full system of equations and its analytic solutions are listed in Appendix \ref{ap:3layer}.

Appendix \ref{ap:shortwaves} explains how the shortwave flux terms ($F_3$, $F_2$, $F_1$ and $F_0$) are determined; they also result in analytic expressions that only depend on user-input parameters. Near-immediate feedback while using the application is therefore allowed.

\section{Summary}
The Climate App incorporates simple climate models based on radiative equilibrium. It offers an intuitive and user-friendly way to explain various climate concepts to students in high school and college. Example modules at various stages of development and beta testing include: the Faint Young Sun Paradox, Snowball Earth, Global Warming, and the habitability of the TRAPPIST-1 planets.  If you would like to test activities or have questions about the Climate App then please contact Nick Cowan.\\

\noindent \textbf{Acknowledgements:} The authors thank Dorian Abbot and Tim Merlis for constructive feedback.



\bibliography{apssamp}

\newpage
\onecolumngrid

\appendix
\section{\label{ap:3layer}Three-Layer Radiative Equilibrium}
The system of equations that ensure radiative equilibrium at each atmospheric layer and at the surface is:
\begin{align*}
    2\epsilon_{\rm LW3}\sigma T_3^4&= F_3 + \epsilon_{\rm LW3}\Big(\sigma T_0^4(1-\epsilon_{\rm LW1})(1-\epsilon_{\rm LW2}) + \sigma T_1^4\epsilon_{\rm LW1}(1-\epsilon_{\rm LW2})  + \sigma T_2^4\epsilon_{\rm LW2}\Big)\\
    2\epsilon_{\rm LW2}\sigma T_2^4 &= F_2 + \epsilon_{\rm LW2}\Big(\sigma T_0^4(1-\epsilon_{\rm LW1}) + \sigma T_1^4\epsilon_{\rm LW1} + \sigma T_3^4\epsilon_{\rm LW3}\Big)\\
    2\epsilon_{\rm LW1}\sigma T_1^4&= F_1 + \epsilon_{\rm LW1}\Big(\sigma T_0^4 + \sigma T_2^4\epsilon_{\rm LW2} + \sigma T_3^4\epsilon_{\rm LW3}(1-\epsilon_{\rm LW2})\Big)\\
    \sigma T_0^4&= F_0 + \Big(\sigma T_1^4\epsilon_{\rm LW1} + \sigma T_2^4\epsilon_{\rm LW2}(1-\epsilon_{\rm LW1}) + \sigma T_3^4\epsilon_{\rm LW3}(1-\epsilon_{\rm LW2})(1-\epsilon_{\rm LW1})\Big),
\end{align*}
where $\epsilon_{\rm LW1}$, $\epsilon_{\rm LW2}$ and $\epsilon_{\rm LW3}$ are the longwave emissivity of the three atmopsheric layers (recall that we assume the surface has a longwave emissivity of 1). Solving this system of equation gives:
\small
\begin{align*}
    T_3^4 = &\frac{\epsilon_{\rm LW3}F_1 + \epsilon_{\rm LW3}F_2 + \epsilon_{\rm LW3}F_0 + F_3}{\sigma\epsilon_{\rm LW3}(2-\epsilon_{\rm LW3})}\\
    &\\
    T_2^4 =& \frac{%
   \three( \splitdfrac{-\epsilon_{\rm LW2}2\epsilon_{\rm LW3}F_1 - \epsilon_{\rm LW2}2\epsilon_{\rm LW3}F_2 - \epsilon_{\rm LW2}2\epsilon_{\rm LW3}F_0 - \epsilon_{\rm LW2}2F_3}{%
    \splitdfrac{+\epsilon_{\rm LW2}\epsilon_{\rm LW3}F_1 + 2\epsilon_{\rm LW2}\epsilon_{\rm LW3}F_2 + \epsilon_{\rm LW2}\epsilon_{\rm LW3}F_0 + 2\epsilon_{\rm LW2}F_1}{%
    +2\epsilon_{\rm LW2}F_3 + 2\epsilon_{\rm LW2}F_0 - \epsilon_{\rm LW3}F_2 + 2F_2 }}\three)}%
    {\sigma\epsilon_{\rm LW2}(\epsilon_{\rm LW2}\epsilon_{\rm LW3} - 2\epsilon_{\rm LW2} - 2\epsilon_{\rm LW3} + 4)}\\
    &\\
    T_1^4=&\frac{%
    \raisebox{-25pt}{\seven(} \splitdfrac{-2\epsilon_{\rm LW1}2\epsilon_{\rm LW2}\epsilon_{\rm LW3}F_1 -                \epsilon_{\rm LW1}2\epsilon_{\rm LW2}\epsilon_{\rm LW3}l_ 2 - 2\epsilon_{\rm LW1}2\epsilon_{\rm LW2}\epsilon_{\rm LW3}F_0}{%
    \splitdfrac{+2\epsilon_{\rm LW1}2\epsilon_{\rm LW2}F_1 - \epsilon_{\rm LW1}2\epsilon_{\rm LW2}F_3 + 2\epsilon_{\rm LW1}2\epsilon_{\rm LW2}F_0 + 2\epsilon_{\rm LW1}2\epsilon_{\rm LW3}F_1}{%
    \splitdfrac{+\epsilon_{\rm LW1}2\epsilon_{\rm LW3}F_2 + 2\epsilon_{\rm LW1}2\epsilon_{\rm LW3}F_0 + 2\epsilon_{\rm LW1}2F_2 + 2\epsilon_{\rm LW1}2F_3}{%
    \splitdfrac{+4\epsilon_{\rm LW1}\epsilon_{\rm LW2}\epsilon_{\rm LW3}F_1 + 2\epsilon_{\rm LW1}\epsilon_{\rm LW2}\epsilon_{\rm LW3}F_2 + 3\epsilon_{\rm LW1}\epsilon_{\rm LW2}\epsilon_{\rm LW3}F_0 }{%
    \splitdfrac{-4\epsilon_{\rm LW1}\epsilon_{\rm LW2}F_1 + 2\epsilon_{\rm LW1}\epsilon_{\rm LW2}F_3 - 2\epsilon_{\rm LW1}\epsilon_{\rm LW2}F_0 - 4\epsilon_{\rm LW1}\epsilon_{\rm LW3}F_1}{%
    \splitdfrac{-2\epsilon_{\rm LW1}\epsilon_{\rm LW3}F_2 - 2\epsilon_{\rm LW1}\epsilon_{\rm LW3}F_0 - 4\epsilon_{\rm LW1}F_2 - 4\epsilon_{\rm LW1}F_3}{%
    -4\epsilon_{\rm LW1}F_0 - \epsilon_{\rm LW2}\epsilon_{\rm LW3}F_1 + 2\epsilon_{\rm LW2}F_1 + 2\epsilon_{\rm LW3}F_1 -4F_1}}}}}}\raisebox{-25pt}{\seven)}}%
    {%
    \sigma\epsilon_{\rm LW1}(\epsilon_{\rm LW1}\epsilon_{\rm LW2}\epsilon_{\rm LW3} - 2\epsilon_{\rm LW1}\epsilon_{\rm LW2} -2\epsilon_{\rm LW1}\epsilon_{\rm LW3} + 4\epsilon_{\rm LW1} -2\epsilon_{\rm LW2}\epsilon_{\rm LW3} + 4\epsilon_{\rm LW2} + 4\epsilon_{\rm LW3} - 8)
    }\\
    &\\
    T_0^4 =&\frac{%
    \raisebox{-9pt}{\four(}\splitdfrac{-2\epsilon_{\rm LW1}\epsilon_{\rm LW2}\epsilon_{\rm LW3}F_1 - \epsilon_{\rm LW1}\epsilon_{\rm LW2}\epsilon_{\rm LW3}F_2 - 2\epsilon_{\rm LW1}\epsilon_{\rm LW2}\epsilon_{\rm LW3}F_0 +2\epsilon_{\rm LW1}\epsilon_{\rm LW2}F_1}{%
    \splitdfrac{-\epsilon_{\rm LW1}\epsilon_{\rm LW2}F_3 + 2\epsilon_{\rm LW1}\epsilon_{\rm LW2}F_0 + 2\epsilon_{\rm LW1}\epsilon_{\rm LW3}F_1 + \epsilon_{\rm LW1}\epsilon_{\rm LW3}F_2 +2\epsilon_{\rm LW1}\epsilon_{\rm LW3}F_0}{%
    \splitdfrac{+ 2\epsilon_{\rm LW1}F_2 + 2\epsilon_{\rm LW1}F_3 + 3\epsilon_{\rm LW2}\epsilon_{\rm LW3}F_1 + 2\epsilon_{\rm LW2}\epsilon_{\rm LW3}F_2 +2\epsilon_{\rm LW2}\epsilon_{\rm LW3}F_0 - 2\epsilon_{\rm LW2}F_1}{%
    +2\epsilon_{\rm LW2}F_3 - 2\epsilon_{\rm LW3}F_1 - 2\epsilon_{\rm LW3}F_2 - 4F_1 - 4F_2 - 4F_3 - 8F_0}}}\raisebox{-9pt}{\four)}}%
    {\sigma(\epsilon_{\rm LW1}\epsilon_{\rm LW2}\epsilon_{\rm LW3} - 2\epsilon_{\rm LW1}\epsilon_{\rm LW2} -2\epsilon_{\rm LW1}\epsilon_{\rm LW3}+ 4\epsilon_{\rm LW1} -2\epsilon_{\rm LW2}\epsilon_{\rm LW3}+4\epsilon_{\rm LW2} +4\epsilon_{\rm LW3} - 8)}.
\end{align*}
\normalsize
\newpage
\section{\label{ap:shortwaves} Shortwave Flux Absorption Terms}

Substituting the expressions for $d_i$ and $u_i$ into the equations for $F_i$ yields
\small
\begin{align*}
    &F_3 = \frac{%
    \raisebox{-73pt}{\fourteen(}\splitdfrac{I\cdot \epsilon_\mathrm{SW3}\cdot(1-\alpha_3)\Big(\big(A\alpha_1^2\alpha_2^2\alpha_3\epsilon_\mathrm{SW1}^2\epsilon_\mathrm{SW2}^2\epsilon_\mathrm{SW3} - 4A\alpha_1^2\alpha_2\epsilon_\mathrm{SW1}^2\epsilon_\mathrm{SW2}}{%
    \hspace*{-.5cm}\splitdfrac{-A\alpha_1^2\alpha_3\epsilon_\mathrm{SW1}^2\epsilon_\mathrm{SW3}(\alpha_2\epsilon_\mathrm{SW2} - 2\epsilon_\mathrm{SW2} + 2)^2 - 4A\alpha_1\alpha_2\alpha_3\epsilon_\mathrm{SW1}\epsilon_\mathrm{SW2}\epsilon_\mathrm{SW3} + 16A\alpha_1\epsilon_\mathrm{SW1}}{%
    \hspace*{-.5cm}\splitdfrac{-A\alpha_2^2\alpha_3\epsilon_\mathrm{SW2}^2\epsilon_\mathrm{SW3}(\alpha_1\epsilon_\mathrm{SW1} - 2\epsilon_\mathrm{SW1} + 2)^2 + 4A\alpha_2\epsilon_\mathrm{SW2}(\alpha_1\epsilon_\mathrm{SW1} - 2\epsilon_\mathrm{SW1} + 2)^2}{%
    \splitdfrac{+A\alpha_3\epsilon_\mathrm{SW3}(\alpha_1\epsilon_\mathrm{SW1} - 2\epsilon_\mathrm{SW1} + 2)^2(\alpha_2\epsilon_\mathrm{SW2} - 2\epsilon_\mathrm{SW2} + 2)^2 - 2\alpha_1\alpha_2^2\alpha_3\epsilon_\mathrm{SW1}\epsilon_\mathrm{SW2}^2\epsilon_\mathrm{SW3}}{%
    \hspace*{-.4cm}\splitdfrac{+8\alpha_1\alpha_2\epsilon_\mathrm{SW1}\epsilon_\mathrm{SW2} +
    2\alpha_1\alpha_3\epsilon_\mathrm{SW1}\epsilon_\mathrm{SW3}(\alpha_2\epsilon_\mathrm{SW2} - 2\epsilon_\mathrm{SW2} + 2)^2 + 8\alpha_2\alpha_3\epsilon_\mathrm{SW2}\epsilon_\mathrm{SW3} - 32\big)}{%
    \hspace*{-.4cm}\splitdfrac{-(\alpha_3\epsilon_\mathrm{SW3} - 2\epsilon_\mathrm{SW3} + 2)
    \big(A\alpha_1^2\alpha_2^2\epsilon_\mathrm{SW1}^2\epsilon_\mathrm{SW2}^2 - A\alpha_1^2\alpha_2\epsilon_\mathrm{SW1}^2\epsilon_\mathrm{SW2}(\alpha_2\epsilon_\mathrm{SW2} - 2\epsilon_\mathrm{SW2} + 2)}{%
    \hspace*{-.3cm}\splitdfrac{-2A\alpha_1^2\alpha_2\epsilon_\mathrm{SW1}^2\epsilon_\mathrm{SW2} +
    2A\alpha_1^2\epsilon_\mathrm{SW1}^2\epsilon_\mathrm{SW2}(\alpha_2\epsilon_\mathrm{SW2} - 2\epsilon_\mathrm{SW2} + 2) + 4A\alpha_1^2\epsilon_\mathrm{SW1}^2\epsilon_\mathrm{SW2}}{%
    \hspace*{-.4cm}\splitdfrac{- 4A\alpha_1^2\epsilon_\mathrm{SW1}^2 - 4A\alpha_1\alpha_2\epsilon_\mathrm{SW1}\epsilon_\mathrm{SW2} +
    4A\alpha_1\epsilon_\mathrm{SW1}(\alpha_1\epsilon_\mathrm{SW1} - 2\epsilon_\mathrm{SW1} + 2) + 8A\alpha_1\epsilon_\mathrm{SW1} }{%
    \hspace*{-.4cm}\splitdfrac{-A\alpha_2^2\epsilon_\mathrm{SW2}^2(\alpha_1\epsilon_\mathrm{SW1} - 2\epsilon_\mathrm{SW1} + 2)^2 +
    A\alpha_2\epsilon_\mathrm{SW2}(\alpha_1\epsilon_\mathrm{SW1} - 2\epsilon_\mathrm{SW1} + 2)^2(\alpha_2\epsilon_\mathrm{SW2} - 2\epsilon_\mathrm{SW2} + 2)}{%
    \hspace*{-1cm}\splitdfrac{+2A\alpha_2\epsilon_\mathrm{SW2}(\alpha_1\epsilon_\mathrm{SW1} - 2\epsilon_\mathrm{SW1} + 2)^2 - 
    8A\epsilon_\mathrm{SW1}(\alpha_1\epsilon_\mathrm{SW1} - 2\epsilon_\mathrm{SW1} + 2) -16A\epsilon_\mathrm{SW1}}{%
    \hspace*{-.4cm}\splitdfrac{-2A\epsilon_\mathrm{SW2}(\alpha_1\epsilon_\mathrm{SW1} - 2\epsilon_\mathrm{SW1} + 2)^2(\alpha_2\epsilon_\mathrm{SW2} - 2\epsilon_\mathrm{SW2} + 2) - 
    4A\epsilon_\mathrm{SW2}(\alpha_1\epsilon_\mathrm{SW1} - 2\epsilon_\mathrm{SW1} + 2)^2 }{%
    \splitdfrac{+16A-2\alpha_1\alpha_2^2\epsilon_\mathrm{SW1}\epsilon_\mathrm{SW2}^2 + 2\alpha_1\alpha_2\epsilon_\mathrm{SW1}\epsilon_\mathrm{SW2}(\alpha_2\epsilon_\mathrm{SW2} - 2\epsilon_\mathrm{SW2} + 2)}{%
    \splitdfrac{+4\alpha_1\alpha_2\epsilon_\mathrm{SW1}\epsilon_\mathrm{SW2} -
    4\alpha_1\epsilon_\mathrm{SW1}\epsilon_\mathrm{SW2}(\alpha_2\epsilon_\mathrm{SW2} - 2\epsilon_\mathrm{SW2} + 2)}{%
    -8\alpha_1\epsilon_\mathrm{SW1}\epsilon_\mathrm{SW2} + 8\alpha_1\epsilon_\mathrm{SW1} + 8\alpha_2\epsilon_\mathrm{SW2}\big)\Big)}}}}}}}}}}}}}\hspace{-.6cm}\raisebox{-73pt}{\fourteen)}}{%
    \raisebox{-15pt}{\five(}\splitdfrac{A\alpha_1^2\alpha_2^2\alpha_3\epsilon_{\rm SW1}^2\epsilon_{\rm SW2}^2\epsilon_{\rm SW3} - 4A\alpha_1^2\alpha_2\epsilon_{\rm SW1}^2\epsilon_{\rm SW2} - A\alpha_1^2\alpha_3\epsilon_{\rm SW1}^2\epsilon_{\rm SW3}(\alpha_2\epsilon_{\rm SW2} - 2\epsilon_{\rm SW2} + 2)^2}{%
    \splitdfrac{-4A\alpha_1\alpha_2\alpha_3\epsilon_{\rm SW1}\epsilon_{\rm SW2}\epsilon_{\rm SW3} + 16A\alpha_1\epsilon_{\rm SW1} - A\alpha_2^2\alpha_3\epsilon_{\rm SW2}^2\epsilon_{\rm SW3}(\alpha_1\epsilon_{\rm SW1} - 2\epsilon_{\rm SW1} + 2)^2}{%
    \splitdfrac{+4A\alpha_2\epsilon_{\rm SW2}(\alpha_1\epsilon_{\rm SW1} - 2\epsilon_{\rm SW1} + 2)^2 + A\alpha_3\epsilon_{\rm SW3}(\alpha_1\epsilon_{\rm SW1} - 2\epsilon_{\rm SW1} + 2)^2}{%
    \splitdfrac{\cdot(\alpha_2\epsilon_{\rm SW2} - 2\epsilon_{\rm SW2} + 2)^2-2\alpha_1\alpha_2^2\alpha_3\epsilon_{\rm SW1}\epsilon_{\rm SW2}^2\epsilon_{\rm SW3} +8\alpha_1\alpha_2\epsilon_{\rm SW1}\epsilon_{\rm SW2} }{%
    + 2\alpha_1\alpha_3\epsilon_{\rm SW1}\epsilon_{\rm SW3}(\alpha_2\epsilon_{\rm SW2} - 2\epsilon_{\rm SW2} + 2)^2+8\alpha_2\alpha_3\epsilon_{\rm SW2}\epsilon_{\rm SW3} - 32}}}}\raisebox{-15pt}{\five)}}
\end{align*}
\begin{align*}
    &F_2 = \frac{%
    \raisebox{-15pt}{\five(} \splitdfrac{4I\cdot\epsilon_{\rm SW2}(\alpha_3\epsilon_{\rm SW3} - 2\epsilon_{\rm SW3} + 2)[A\alpha_1^2\epsilon_{\rm SW1}^2\epsilon_{\rm SW2}(\alpha_2 - 1)}{%
    \splitdfrac{-A\alpha_1^2\epsilon_{\rm SW1}^2(\alpha_2 - 1) + A\alpha_1\epsilon_{\rm SW1}(\alpha_2 - 1)(\alpha_1\epsilon_{\rm SW1} - 2\epsilon_{\rm SW1} + 2)}{%
    \splitdfrac{-2A\epsilon_{\rm SW1}(\alpha_2 - 1)(\alpha_1\epsilon_{\rm SW1} - 2\epsilon_{\rm SW1} + 2) - 4A\epsilon_{\rm SW1}(\alpha_2 - 1)}{%
    \splitdfrac{-A\epsilon_{\rm SW2}(\alpha_2 - 1)(\alpha_1\epsilon_{\rm SW1} - 2\epsilon_{\rm SW1} + 2)^2 + 4A(\alpha_2 - 1)}{-2\alpha_1\epsilon_{\rm SW1}\epsilon_{\rm SW2}(\alpha_2 - 1) + 2\alpha_1\epsilon_{\rm SW1}(\alpha_2 - 1) + 4\alpha_2 - 4]}}}}\raisebox{-15pt}{\five)}}
    {%
    \raisebox{-15pt}{\five(}\splitdfrac{A\alpha_1^2\alpha_2^2\alpha_3\epsilon_{\rm SW1}^2\epsilon_{\rm SW2}^2\epsilon_{\rm SW3} - 4A\alpha_1^2\alpha_2\epsilon_{\rm SW1}^2\epsilon_{\rm SW2} - A\alpha_1^2\alpha_3\epsilon_{\rm SW1}^2\epsilon_{\rm SW3}(\alpha_2\epsilon_{\rm SW2} - 2\epsilon_{\rm SW2} + 2)^2}{%
    \splitdfrac{-4A\alpha_1\alpha_2\alpha_3\epsilon_{\rm SW1}\epsilon_{\rm SW2}\epsilon_{\rm SW3} + 16A\alpha_1\epsilon_{\rm SW1} - A\alpha_2^2\alpha_3\epsilon_{\rm SW2}^2\epsilon_{\rm SW3}(\alpha_1\epsilon_{\rm SW1} - 2\epsilon_{\rm SW1} + 2)^2}{%
    \splitdfrac{+4A\alpha_2\epsilon_{\rm SW2}(\alpha_1\epsilon_{\rm SW1} - 2\epsilon_{\rm SW1} + 2)^2 + A\alpha_3\epsilon_{\rm SW3}(\alpha_1\epsilon_{\rm SW1} - 2\epsilon_{\rm SW1} + 2)^2}{%
    \splitdfrac{\cdot(\alpha_2\epsilon_{\rm SW2} - 2\epsilon_{\rm SW2} + 2)^2-2\alpha_1\alpha_2^2\alpha_3\epsilon_{\rm SW1}\epsilon_{\rm SW2}^2\epsilon_{\rm SW3} +8\alpha_1\alpha_2\epsilon_{\rm SW1}\epsilon_{\rm SW2} }{%
    + 2\alpha_1\alpha_3\epsilon_{\rm SW1}\epsilon_{\rm SW3}(\alpha_2\epsilon_{\rm SW2} - 2\epsilon_{\rm SW2} +  2)^2+8\alpha_2\alpha_3\epsilon_{\rm SW2}\epsilon_{\rm SW3} - 32}}}}\raisebox{-15pt}{\five)}}
\end{align*}
\begin{align*}
    &F_1 = \frac{%
     \raisebox{-9pt}{\four(}\splitdfrac{4I\cdot\epsilon_{\rm SW1}(\alpha_3\epsilon_{\rm SW3} - 2\epsilon_{\rm SW3} + 2)[-A\alpha_1\alpha_2\epsilon_{\rm SW1}\epsilon_{\rm SW2}(\alpha_1 - 1)}{%
    \splitdfrac{+2A\alpha_1\epsilon_{\rm SW1}\epsilon_{\rm SW2}(\alpha_1 - 1) + A\alpha_2\epsilon_{\rm SW2}(\alpha_1 - 1)(\alpha_1\epsilon_{\rm SW1} - 2\epsilon_{\rm SW1} + 2)}{%
    \splitdfrac{-4A\epsilon_{\rm SW1}(\alpha_1 - 1) - 2A\epsilon_{\rm SW2}(\alpha_1 - 1)(\alpha_1\epsilon_{\rm SW1} - 2\epsilon_{\rm SW1} + 2)}{%
    +4A(\alpha_1 - 1) + 4\alpha_1 + 2\alpha_2\epsilon_{\rm SW2}(\alpha_1 - 1) - 4\epsilon_{\rm SW2}(\alpha_1 - 1) - 4]}}}\raisebox{-9pt}{\four)}}{%
    \raisebox{-15pt}{\five(}\splitdfrac{A\alpha_1^2\alpha_2^2\alpha_3\epsilon_{\rm SW1}^2\epsilon_{\rm SW2}^2\epsilon_{\rm SW3} - 4A\alpha_1^2\alpha_2\epsilon_{\rm SW1}^2\epsilon_{\rm SW2} - A\alpha_1^2\alpha_3\epsilon_{\rm SW1}^2\epsilon_{\rm SW3}(\alpha_2\epsilon_{\rm SW2} - 2\epsilon_{\rm SW2} + 2)^2}{%
    \splitdfrac{-4A\alpha_1\alpha_2\alpha_3\epsilon_{\rm SW1}\epsilon_{\rm SW2}\epsilon_{\rm SW3} + 16A\alpha_1\epsilon_{\rm SW1} - A\alpha_2^2\alpha_3\epsilon_{\rm SW2}^2\epsilon_{\rm SW3}(\alpha_1\epsilon_{\rm SW1} - 2\epsilon_{\rm SW1} + 2)^2}{%
    \splitdfrac{+4A\alpha_2\epsilon_{\rm SW2}(\alpha_1\epsilon_{\rm SW1} - 2\epsilon_{\rm SW1} + 2)^2 + A\alpha_3\epsilon_{\rm SW3}(\alpha_1\epsilon_{\rm SW1} - 2\epsilon_{\rm SW1} + 2)^2}{%
    \splitdfrac{\cdot(\alpha_2\epsilon_{\rm SW2} - 2\epsilon_{\rm SW2} + 2)^2-2\alpha_1\alpha_2^2\alpha_3\epsilon_{\rm SW1}\epsilon_{\rm SW2}^2\epsilon_{\rm SW3} +8\alpha_1\alpha_2\epsilon_{\rm SW1}\epsilon_{\rm SW2} }{%
    + 2\alpha_1\alpha_3\epsilon_{\rm SW1}\epsilon_{\rm SW3}(\alpha_2\epsilon_{\rm SW2} - 2\epsilon_{\rm SW2} + 2)^2+8\alpha_2\alpha_3\epsilon_{\rm SW2}\epsilon_{\rm SW3} - 32}}}}\raisebox{-15pt}{\five)}}\\
    &\\
    &F_0 = \frac{%
    \two(\splitdfrac{%
    4I\cdot(\alpha_3\epsilon_{\rm SW3} - 2\epsilon_{\rm SW3} + 2)[4A + 2\alpha_1\epsilon_{\rm SW1}(A - 1) +\alpha_2\epsilon_{\rm SW2}(A - 1)(\alpha_1\epsilon_{\rm SW1} - 2\epsilon_{\rm SW1} + 2)}{%
    -4\epsilon_{\rm SW1}(A - 1) - 2\epsilon_{\rm SW2}(A - 1)(\alpha_1\epsilon_{\rm SW1} - 2\epsilon_{\rm SW1} + 2) - 4]}\two)}%
    {%
    \raisebox{-15pt}{\five(}\splitdfrac{A\alpha_1^2\alpha_2^2\alpha_3\epsilon_{\rm SW1}^2\epsilon_{\rm SW2}^2\epsilon_{\rm SW3} - 4A\alpha_1^2\alpha_2\epsilon_{\rm SW1}^2\epsilon_{\rm SW2} -A\alpha_1^2\alpha_3\epsilon_{\rm SW1}^2\epsilon_{\rm SW3}(\alpha_2\epsilon_{\rm SW2} - 2\epsilon_{\rm SW2} + 2)^2}{%
    \splitdfrac{-4A\alpha_1\alpha_2\alpha_3\epsilon_{\rm SW1}\epsilon_{\rm SW2}\epsilon_{\rm SW3} + 16A\alpha_1\epsilon_{\rm SW1} - A\alpha_2^2\alpha_3\epsilon_{\rm SW2}^2\epsilon_{\rm SW3}(\alpha_1\epsilon_{\rm SW1} - 2\epsilon_{\rm SW1} + 2)^2 }{%
    \splitdfrac{+4A\alpha_2\epsilon_{\rm SW2}(\alpha_1\epsilon_{\rm SW1} - 2\epsilon_{\rm SW1} + 2)^2+A\alpha_3\epsilon_{\rm SW3}(\alpha_1\epsilon_{\rm SW1} - 2\epsilon_{\rm SW1} + 2)^2 }{\splitdfrac{\cdot(\alpha_2\epsilon_{\rm SW2} - 2\epsilon_{\rm SW2} + 2)^2- 2\alpha_1\alpha_2^2\alpha_3\epsilon_{\rm SW1}\epsilon_{\rm SW2}^2\epsilon_{\rm SW3}+8\alpha_1\alpha_2\epsilon_{\rm SW1}\epsilon_{\rm SW2} }{+2\alpha_1\alpha_3\epsilon_{\rm SW1}\epsilon_{\rm SW3}(\alpha_2\epsilon_{\rm SW2}- 2\epsilon_{\rm SW2} + 2)^2 + 8\alpha_2\alpha_3\epsilon_{\rm SW2}\epsilon_{\rm SW3} - 32}}}}\raisebox{-15pt}{\five)}}.
\end{align*}


\end{document}